\documentclass[11pt,twoside]{article}

\usepackage{asp2010}
\usepackage{lscape}
\usepackage{natbib}
\usepackage{epsf}
\usepackage{graphicx}
\begin{document}

\title{Simulations of Dense Stellar Systems
       with the AMUSE Software Toolkit}

\author{Stephen McMillan,$^{1}$
        Simon Portegies Zwart,$^{2}$
        Arjen van Elteren,$^{2}$
        Alfred Whitehead$^{1}$}

\affil{
$^1$Physics Department, Drexel University, Philadelphia, PA
      19104, USA\\
$^2$Sterrewacht Leiden, P.O. Box 9513, 2300 RA Leiden,
      The Netherlands 
}

\begin{abstract}
  We describe AMUSE, the Astrophysical Multipurpose Software
  Environment, a programming framework designed to manage multi-scale,
  multi-physics simulations in a hierarchical, extensible, and
  internally consistent way.  Constructed as a collection of
  individual modules, AMUSE allows computational tools for different
  physical domains to be easily combined into a single task.  It
  facilitates the coupling of modules written in different languages
  by providing inter-language tools and a standard programming
  interface that represents a balance between generality and
  computational efficiency.  The framework currently incorporates the
  domains of stellar dynamics, stellar evolution, gas dynamics, and
  radiative transfer.  We present some applications of the framework
  and outline plans for future development of the package.
\end{abstract}

\section{Introduction}\label{Sect:Introduction}

Many areas of computational astrophysics entail simultaneous solution
of systems of equations spanning multiple physical domains.  These
domains may themselves span broad ranges in length and time scales,
and separate domains may be tightly coupled on the scales of interest.
The combination of many competing physical processes and large dynamic
range, together with the sheer size of the computation, represents a
major theoretical challenge for computational astrophysics.  As
simulations and data analysis become increasingly complex,
computational scientists face an increasing need for flexible
frameworks capable of integrating new and existing scientific codes
and allowing scientists to easily build their simulation workflow.

The targets of interest here, young massive star clusters and galactic
nuclei, are dense stellar environments in which gravitational
dynamics, radiative processes, stellar physics, and gas dynamics all
play important roles.  Spatial and temporal scales range from
$10^3$\,m and $10^{-3}$\,s on the small end to $10^{20}$\,m and
$10^{17}$\,s at the other extreme.  Close encounters and physical
collisions among stars and binaries are commonplace, and large and
small scales are intimately coupled by stellar mass loss, binary
heating, stellar collisions, dynamical mass segregation, and core
collapse.  The number of stars can exceed $10^6$ in many cases.
Combining all these elements within a large-scale simulation of such a
system poses significant software development problems.

The present generation of cluster simulation packages---often
generically referred to as ``kitchen-sink'' codes---have been very
successful in modeling the long-term dynamical evolution of star
clusters, from a few megayears after formation to their eventual
dissolution possibly gigayears later in the galactic tidal field.
These packages include both N-body
\citep{Aarseth2003,PortegiesZwart_etal2001} and Monte Carlo codes
\citep{FregeauRasio2007,Giersz_etal2008}.  All contain sophisticated
and comprehensive treatments of stellar dynamics and binary/multiple
interactions---a reflection of their historical development---but
these are generally coupled with much more approximate treatments of
other physical aspects of the system.  For example, stellar evolution
is typically calculated as a look-up from precomputed results
\citep{Hurley_etal2000}, while binary evolution consists of a set of
rules of varying accuracy, implemented on top of the stellar evolution
subsystem.

Other aspects of the simulation, such as stellar collisions, are
treated even more approximately.  Stars are modeled as ``sticky
spheres'' which merge when they come into contact, preserving
virtually none of the underlying stellar physics.  Still other
physical processes, such as global gas dynamics and the effects of
stellar winds and radiation, which are now regarded as critical
components as we push our simulations back into the star-formation
phase, are not included at all.  An additional limitation of most
kitchen sink codes is that specific implementations of each physical
process are hard-coded into the simulation, so adoption of a
particular package implies a particular choice of dynamical
integrator, stellar modeling, binary evolution, and so on.

We submit that the monolithic design and consequent internal
complexity of existing codes are limiting their future development,
making alternate treatments of existing physics hard to implement and
new physical processes even more difficult to deploy.  We expect that
these codes will encounter more and greater structural challenges as
the demands of greater realism and more comprehensive content mount.

In this paper we describe AMUSE, the Astrophysical Multipurpose
Software Environment, a modular simulation framework designed to
address the shortcomings just described.  The individual modules in
AMUSE contain dedicated and efficient implementations of specific
pieces of the calculation, and are linked by a high-level scripting
language to ensure flexibility and facilitate management.  Modules can
contain wrapped legacy code or new code developed specifically for the
project.  All have a standard interface exposing only necessary
functionality, allowing them to be easily mixed and replaced as
needed.  We use the term ``community code'' to refer collectively to
public domain codes encompassed by the AMUSE framework.  Section
\ref{Sect:AMUSE} discusses the structure of AMUSE in more detail;
Section \ref{Sect:Applications} presents some simple applications of
the framework.

\section{The AMUSE Software Framework}\label{Sect:AMUSE}

The global structure of the AMUSE environment is illustrated in Figure
1.  In the AMUSE programming model, each piece of physics (advance the
stellar or hydrodynamics to a specified time, evolve a star, collide
two stars, etc.) is implemented as a module with a standard interface
onto the rest of the system, but the details are private to each
module.  For example, all stellar modules include accessor functions
that provide information on the mass and radius of a specified star,
but the details of what a ``star'' is (an analytic formula, an entry
in a look-up table, or a set of 1- or 2-D arrays describing the run of
density, temperature, composition, etc.)  are internal to the module
and are normally invisible to the user.  Objects within each module
are identified by a global ID, and it is the responsibility of the
module implementation to provide the required accessor functions for a
specified ID.

\begin{figure}[!t]
  \begin{center}
  \includegraphics[width=4.75in, bb = 50 45 750 520,
                   clip = true]{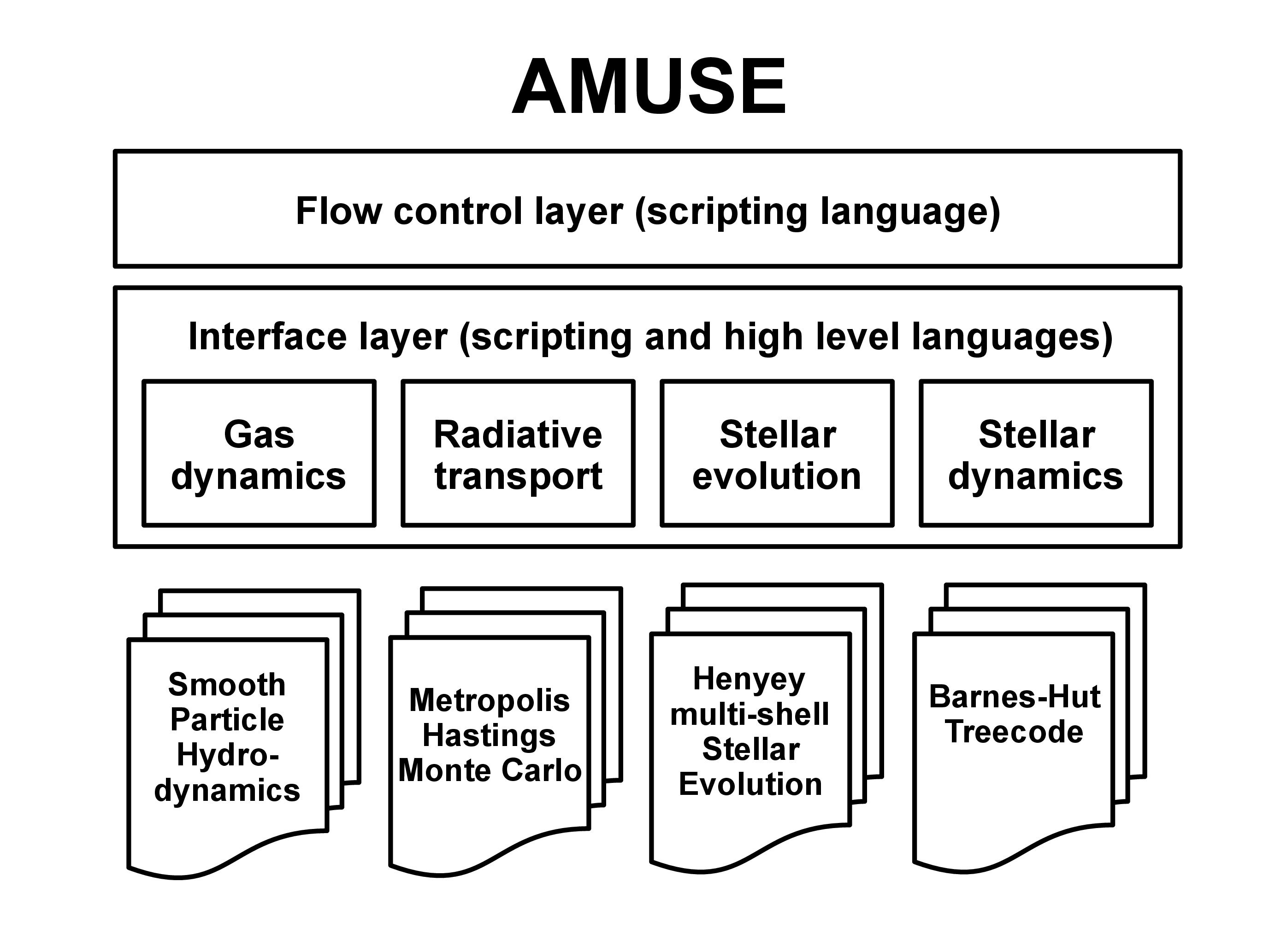}

  \end{center}
  \caption{The AMUSE environment.  The top-level flow control layer is
    typically a custom GUI or user-written Python script that
    specifies the structure of the program, effectively replacing the
    top-level loop of a traditional program.  Each of the four physics
    areas shown in the interface layer may be instantiated by one of
    several modules, allowing arbitrary combinations to be explored.}
\end{figure}

The high-level ``glue'' language for AMUSE is Python, chosen for its
rich feature set, ease of programming and rapid prototyping,
object-oriented capabilities, large user base in the astronomical
community, and extensive user-written software.  The design of AMUSE
places no restrictions on the choice of language for any given module,
except that it must support the parallel Message Passing Interface
(MPI).\footnote{See {\tt http://www.mcs.anl.gov/mpi}.} In a typical
application, the top-level loop (the flow control layer in Figure 1)
of a simulation is written entirely in Python, allowing monitoring,
analysis, graphics, grid management, and other Python tools to be
employed.  The modular design and the use of private internal data
minimizes both the computational overhead of the Python code segments
and data flow between modules.  The relatively low speed of the
language does not significantly impact performance, because in
practice virtually all of the computational load is carried by the
(high-performance) modules.

The concept and value of modular software frameworks for program
integration are familiar to most computational scientists.  Perhaps
less obvious is the use of MPI as the communication tool among
modules.  In the initial implementation of AMUSE, individual modules
written in Fortran 77, Fortran 90, Fortran 95, C, and C++ were
interfaced with Python using f2py or swig.\footnote{See {\tt
    http://www.scipy.org/F2Py} and {\tt http://www.swig.org}.}  This
approach works well for simple demonstration programs
\citep[see][]{PortegiesZwart_etal2009}, but it has a number of serious
technical drawbacks when deployed in a parallel, high-performance
environment.  It (1) imposes namespace restrictions that cause
conflicts between independent modules, (2) makes it impossible to
instantiate multiple independent copies of a given module, and,
perhaps most importantly, (3) rules out incorporation of parallel
modules into the AMUSE framework.

All of these problems can be eliminated by replacing the standard
swig/f2py interface by an explicitly parallel structure in which MPI
is used throughout the AMUSE system for communication between all
modules (serial or parallel).  Like Python, MPI is easily available,
well documented, and widely used in the astronomical community.  As a
practical matter, the process of generating the necessary interface
code is completely automated, using a syntax similar to that found in
swig, so ease of implementation is not a significant consideration for
the applications programmer.  The result is that serial and parallel
modules are indistinguishable one another, as seen from the flow
control level (Figure 1), making them easy to combine, largely
transparently to the user.  In addition, the top-level script can run
modules concurrently, should the structure of the problem allow it.

Currently, AMUSE contains at least two (and in most cases
substantially more) independent modules for each physical process
supported, allowing ``plug and play'' interchangeability between
implementations.  This and encourages approach enables direct
comparison and calibration of different implementations of the same
physical processes, and facilitates experimentation in constructing
new models.  For the current list of available modules and their
properties, and to download and install AMUSE, see the project web
site, {\tt http://amusecode.org}.

Figure 2 illustrates schematically the AMUSE programming environment
in a star cluster simulation typical of those carried out by current
kitchen-sink codes.  The python script acts as a scheduler for the
stellar dynamical and stellar evolution modules and takes appropriate
action when unscheduled events (such as collisions or supernovae)
occur.  In most cases there is no need for communication between
modules beyond the data transferred through the standard interface.
However, additional information may be needed in some
circumstances---for example, the collision and stellar evolution
modules may need to exchange detailed structural data before and after
a physical stellar collision.  At present all such exchanges are
implemented as two MPI transfers (red arrows) through and under the
control of the top-level Python layer.  So far, we have not found this
approach to contribute significantly to the total cost of a
simulation.  Should these data transfers become too expensive, direct
module-to-module communication can be implemented.

\begin{figure}[!ht]
  \begin{center}
  \includegraphics[width=4.75in, bb = 30 60 680 475,
                   clip = true]{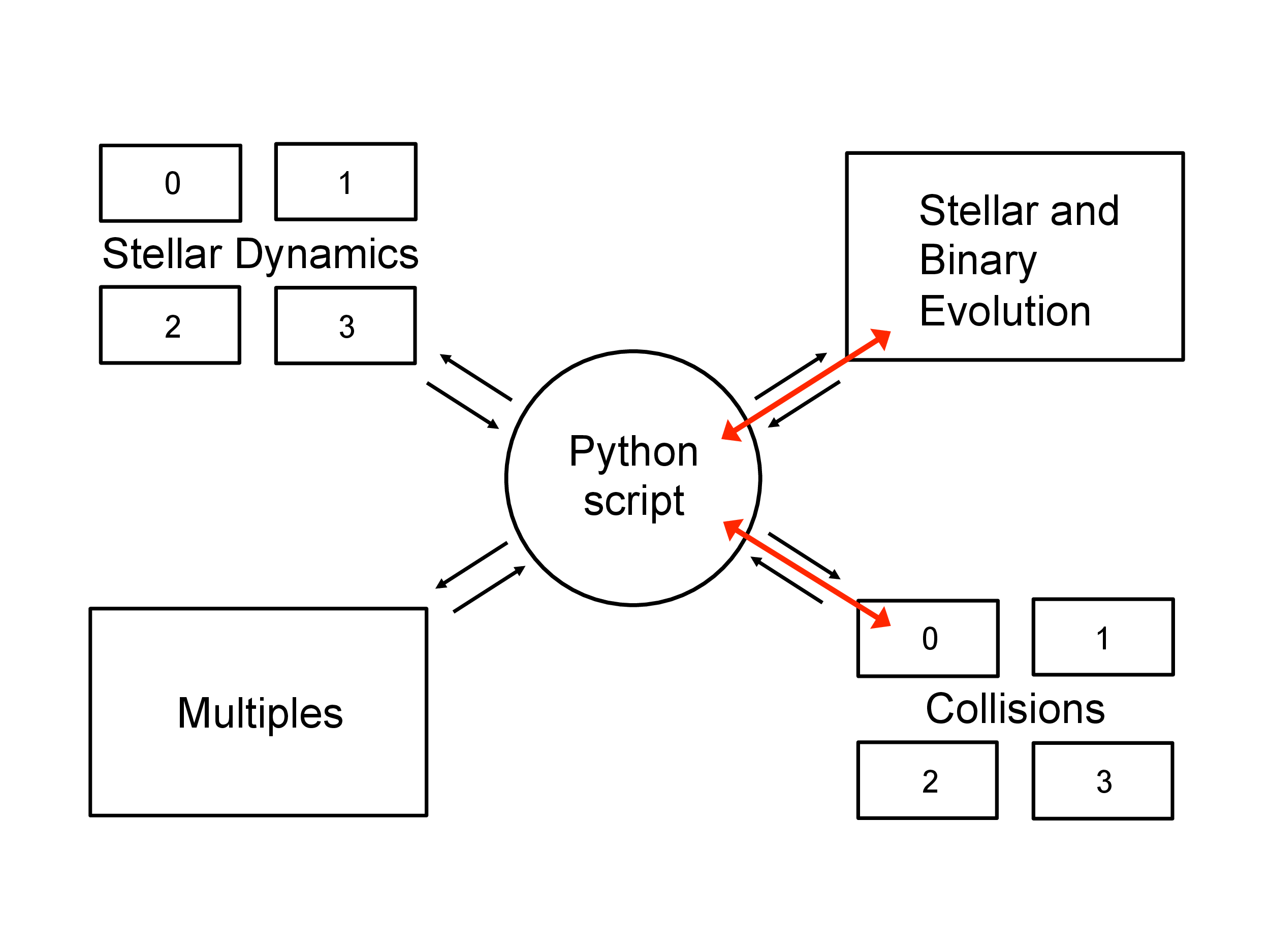}
  \end{center}
  \caption{Structure of a typical AMUSE program.  The user-written
    Python script acts as a scheduler for the modules, and as a
    manager for unscheduled events.  The short (black) arrows indicate
    routine communication through the module interfaces.  Additional
    module to module communication is managed through the Python
    layer, as indicated by the long (red) arrows.  All communications
    are carried out using MPI.  In this case, two of the modules are
    serial, the other two parallel, but the script author need not be
    aware of their internal organization.}
\end{figure}

N-body experts will notice that binary and multiple interactions
(including two-body encounters) are treated by the scheduler at the
same level as actual physical collisions.  In the AMUSE model, the
dynamical integrator follows only the centers of mass of binaries and
multiples, which are regarded as unperturbed until a close encounter
occurs.  Encounters are flagged by the dynamics module, integrated to
completion in isolation by the multiples module, and the products are
then reinserted into the dynamics module.  This approach is unusual
for N-body codes, in which binaries and multiples are generally
integrated simultaneously with the rest of the dynamics, and
represents perhaps the greatest departure of AMUSE from traditional
N-body practice.  (See, however, the ``gorilla'' code described by
\citeauthor{TanikawaFukushige2009} \citeyear{TanikawaFukushige2009}.)
However, this is precisely the way in which multiples are handled in
the leading Monte Carlo codes.  To the best of our knowledge, it has
never been demonstrated that the neglect of weak perturbations on
binaries introduces any bias into the large-scale dynamics of the
system.  Secular evolution of a binary due to occasional wide
encounters can be modeled by integrating the orbit-averaged equations
for the binary elements as the center of mass moves.

\section{Applications of AMUSE}\label{Sect:Applications}

AMUSE has been applied to a number of ``proof of concept'' problems
involving interactions between the large-scale dynamical modules and
the stellar evolution and multiples modules.  We describe two of them
here.  For a more sophisticated simulation carried out using AMUSE,
see the contribution by Portegies Zwart et al. elsewhere in these
proceedings.

Figure 3 is drawn from a study of mass loss from star clusters carried
out by Whitehead et al. \citetext{2012, in preparation}.  The goals of
the investigation are (1) to demonstrate that AMUSE can reproduce
existing results for large simulations, (2) to quantify the run-to-run
variations in simulations differing only in the random seed used to
generate their initial conditions, and (3) to determine the effect of
the choice of stellar evolution model on cluster lifetimes.  All runs
are performed using 16k or 32k particles initialized from
\citet{King1966} model distributions, with power-law ($dN/dM \propto
M^\alpha$) stellar mass spectra, and tidal radii corresponding to one
of the \citet{ChernoffWeinberg1990} families (where increasing family
number indicates larger galactocentric distance).  In all cases, the
AMUSE {\tt ph4} dynamical module is used.  This C++ integrator
includes an MPI parallel fourth-order Hermite integration scheme
\citep{MakinoAarseth1992} with block time steps and GPU acceleration.
The {\tt ph4} module is coupled with one of several stellar evolution
modules: the simple scheme adopted by \citet[CW in the figure, written
in Python]{ChernoffWeinberg1990}, the look-up formula presented by
\citet[EFT89/C]{Eggleton_etal1989}, the {\tt SeBa} package drawn from
Starlab's {\tt kira} integrator
\citep[SeBa/C++]{PortegiesZwart_etal2001}, and the two-dimensional SSE
interpolation scheme of \citet[SSE/Fortran-77]{Hurley_etal2000}.

\begin{figure}[!ht]
  \begin{center}
  \includegraphics[width=2.6in, bb = 80 40 600 470,
                   clip = true]{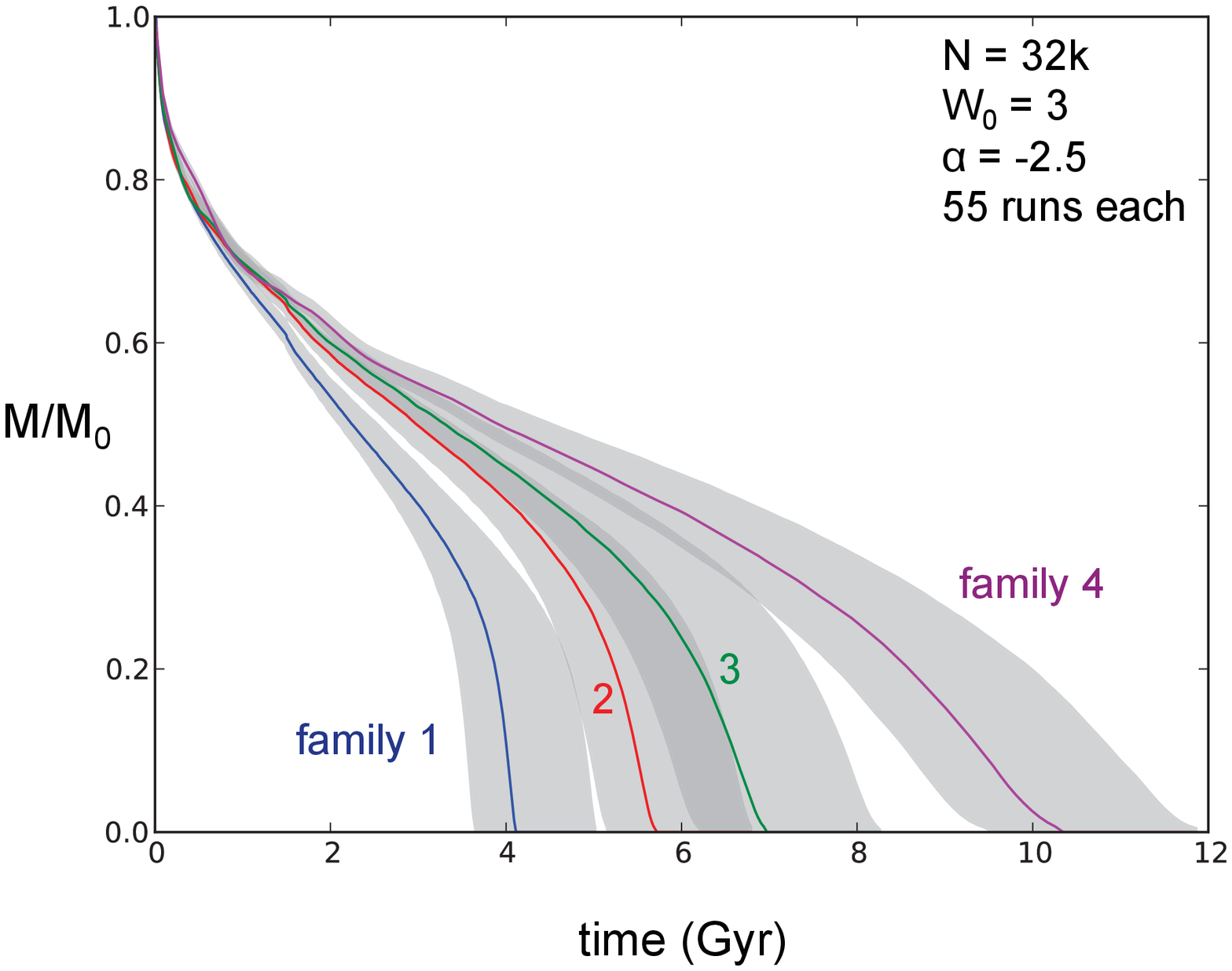}~
  \includegraphics[width=2.6in, bb = 80 40 600 470,
                   clip = true]{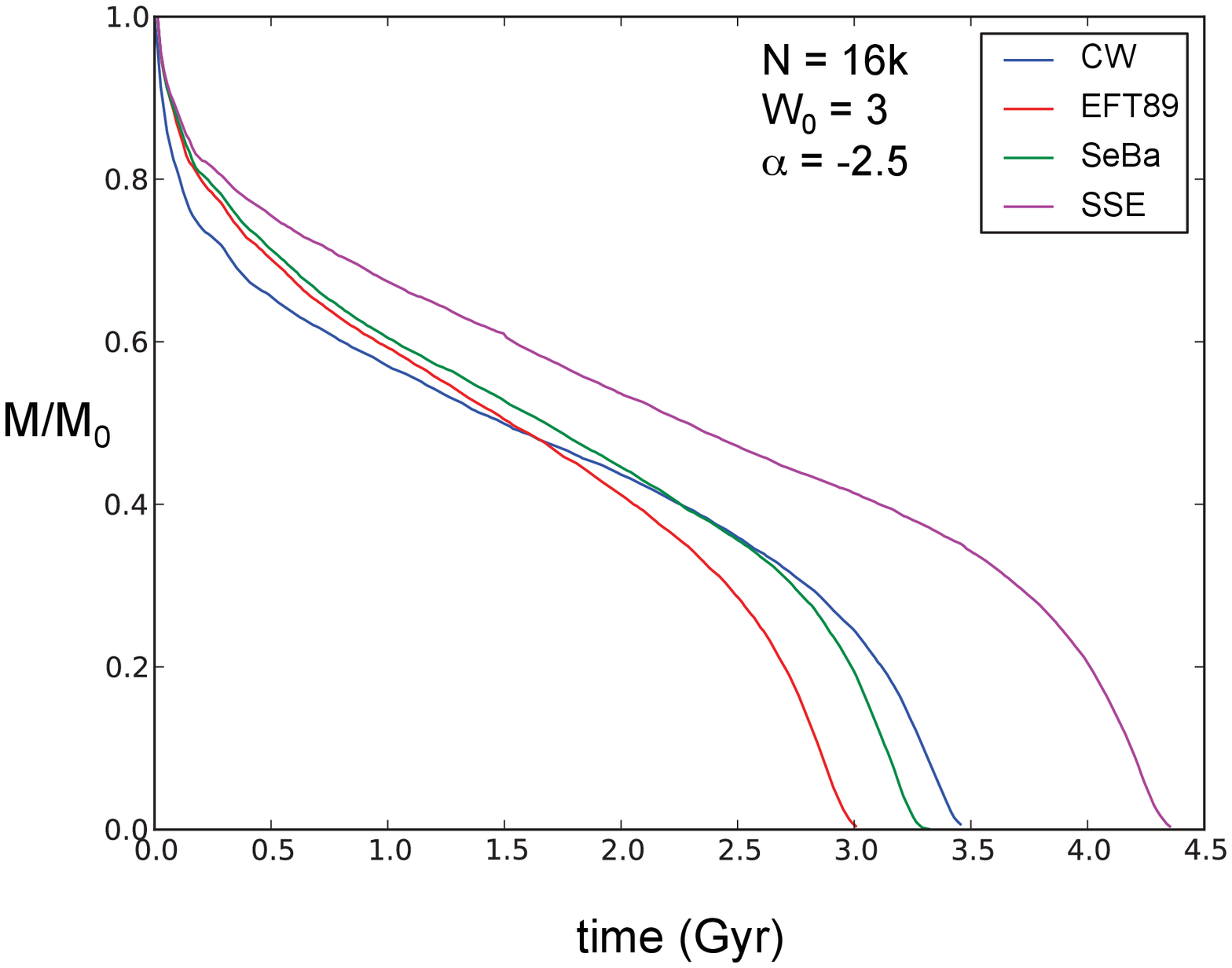}
  \end{center}
  \caption{AMUSE simulations of mass loss in tidally limited star
    clusters.  (a) Comparison of the evolution of different
    \citet{ChernoffWeinberg1990} families.  All models began as
    tidally limited $W_0=3$ King models at one of four galactocentric
    radii, with a power-law $dN/dM\propto M^{-2.5}$ stellar mass
    function.  Solid (colored) lines represent the median behavior of
    55 simulations performed for each family; grey-shaded regions
    represent the full range of the results.  (b) Effect of varying
    the prescription for stellar evolution in otherwise identical
    simulations.  From left to right at the lower right of the figure,
    the curves correspond to the EFT89, SeBa, CW, and SSE stellar
    evolution modules.}
\end{figure}

The results shown in Figure 3(a), using the SeBa stellar evolution
module are in good agreement with the N-body and Fokker--Planck
simulations reported by \citet{TakahashiPortegiesZwart2000}.  We find
similarly good agreement in most of the other cases studied.  We note
in passing that the AMUSE runs are slightly faster than the
corresponding calculations performed using Starlab.  Since the {\tt
  ph4} internals are somewhat similar to those in {\tt kira}, and in
particular employ similar GPU acceleration, we attribute this to the
fact that {\tt ph4} spends less time checking for and handling binary
interactions.  The grey-shaded areas in the figure show the entire
range of results for 55 models initialized from the same system
parameters, allowing us to measure the spread in the numerical
results.  Most of this spread in fact arises from variations in the
initial stellar masses, rather than variations in their positions or
velocities---the mass of the most massive star has a large effect on
the lifetime of the system.

Figure 3(b) shows four runs carried out using identical initial
conditions but with four different stellar evolution modules---a
simple task using AMUSE, entailing changes in just two lines of the
driving script.  We find that the choice of module---not an option
with kitchen-sink codes, and one not normally thought of as a critical
choice---can have a large impact on the lifetime of the system.  In
this example, the AMUSE capacity for easy code comparison provides
valuable insights into the systematic errors inherent in our
calculations.  Such a comparison is not easy to make using traditional
monolithic codes because of the difficulty in implementing even a
simple algorithmic change within those frameworks.

\begin{figure}[!ht]
  \begin{center}
  \includegraphics[width=4in, bb = 30 20 680 490,
                   clip = true]{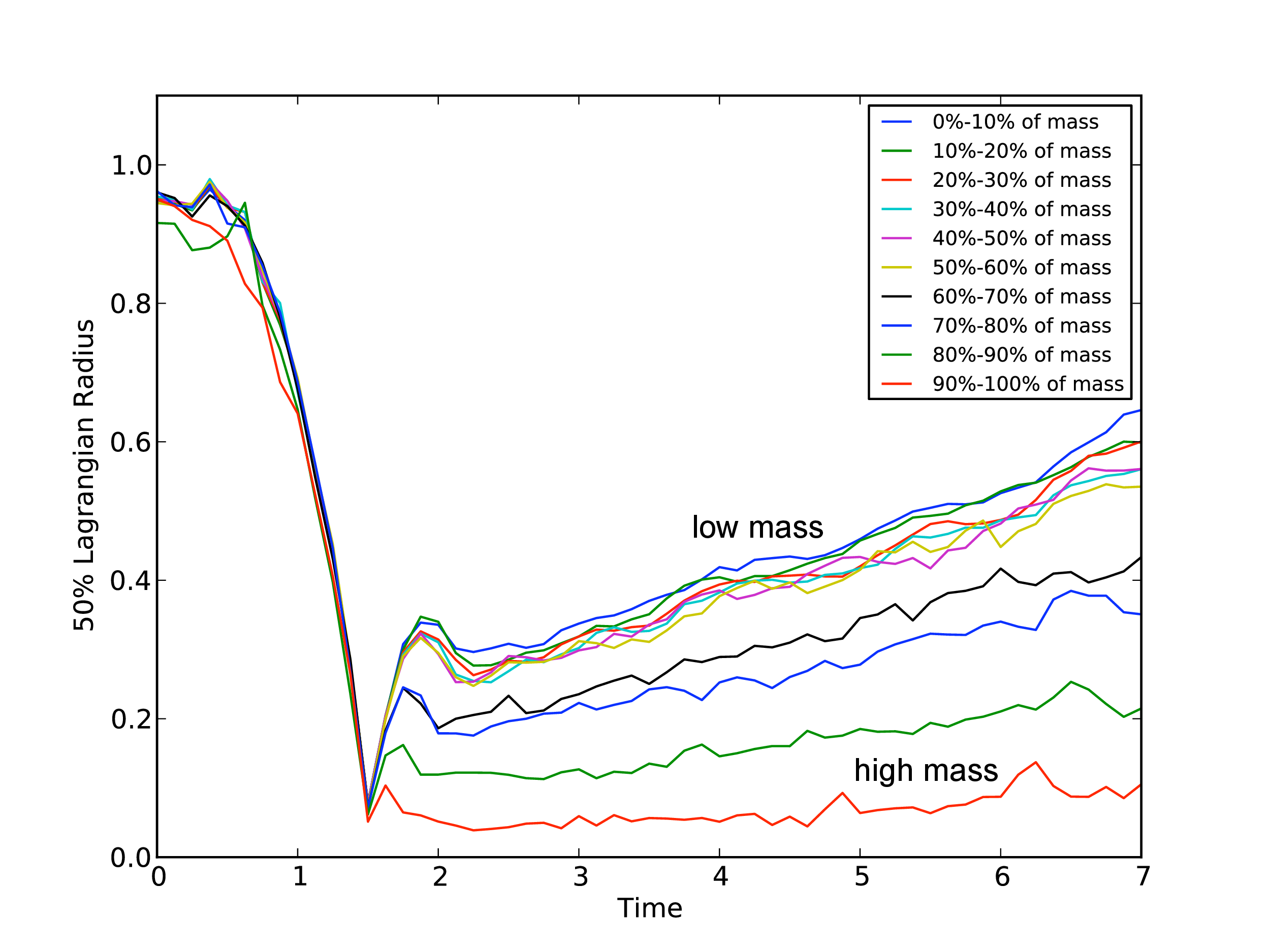}
  \end{center}
  \caption{AMUSE simulation of the collapse of an initially
    homogeneous spherical system with a \citet{Kroupa2002} stellar
    mass function.  The half-mass radii of the indicated mass groups
    are shown.  The bottom four lines after the collapse represent the
    top four mass groups, their half-mass radii decreasing with mass,
    indicating strong mass segregation.}
\end{figure}

Figure 4 shows an AMUSE simulation that provides a stringent test of
how the framework handles binaries and close encounters.  The initial
conditions consist of a cold homogeneous sphere of $N=10^4$ particles
with a \citet{Kroupa2002} mass distribution.  The scientific interest
of this simulation lies in the fact that this system experiences mass
segregation on a dynamical time scale, as was previously noted by
\citet{Vesperini_etal2006,Vesperini_etal2009} and
\citet{Allison_etal2009}.  The segregation is clearly evident in the
figure, which shows the time evolution of the half-mass radii of the
particle sets making up the bottom 10 percent, 10--20 percent, 20--30
percent, etc., of the cumulative mass distribution.  No segregation is
evident before the ``bounce'' at $t\sim1.5$ initial dynamical times,
while immediately afterward the highest mass groups are clearly
ordered by radius.

\citet{Allison_etal2009} suggested that this unexpected result might
be due to enhanced relaxation around the high density bounce, but this
would only be possible if the system were still quite cold at this
time, and our simulations suggest this is not the case.  Instead, it
appears that the system fragments as it collapses, as described by
\citet{Aarseth_etal1988}, and that the fragments mass segregate quite
early in the collapse.  Significant segregation within the clumps is
already established by $t\sim1$, well before the bounce, and is
preserved when the clumps subsequently merge, as described by
\citet{McMillan_etal2007}.

Our simulations test the binary handling abilities of AMUSE because
the deep collapse---by a factor of $N^{1/3}$ or $N^{2/3}$, depending
on the details of the initial conditions---leads to many close
encounters and results in binary formation and interaction among the
most massive stars in the system.  The late expansion of the low-mass
stars is driven by heating due to these binaries.  In side-by-side
tests we find no statistically significant differences between AMUSE
and Starlab, in either the overall behavior of the system (e.g. Figure
4) or in the integration errors incurred.

We interpret these results as encouraging signs that AMUSE has crossed
the threshold where it now incorporates all of the key functionality
found in the leading kitchen-sink codes.  Upcoming development of the
framework will focus on (1) improved handling at the python level of
interactions among community modules, including feedback between the
multiples, collisions, and stellar/binary evolution modules, and (2)
full integration of the new collisional dynamical modules with the
global gas dynamics and radiative transport subsystems.

\acknowledgments
This work was supported by NSF grant AST-0708299 in the U.S. and NWO
(grants \#643.200.503, \#639.073.803 and \#614.061.608), NOVA and the
LKBF in the Netherlands.  Part of the work was done while the authors
visited the Center for Planetary Science (CPS) in Kobe, Japan, during
a visit that was funded by the HPCI Strategic Program of MEXT.  We are
grateful for their hospitality.

\bibliography{McMillan_S}

\begin{thebibliography}{}
\expandafter\ifx\csname natexlab\endcsname\relax\def\natexlab#1{#1}\fi
\expandafter\ifx\csname url\endcsname\relax
  \def\url#1{\texttt{#1}}\fi
\expandafter\ifx\csname urlprefix\endcsname\relax\def\urlprefix{URL }\fi
\providecommand{\eprint}[2][]{\url{#2}}

\bibitem[{{Aarseth}(2003)}]{Aarseth2003}
{Aarseth}, S.~J. 2003, {Gravitational N-Body Simulations} (Cambridge University
  Press)

\bibitem[{{Aarseth} et~al.(1988){Aarseth}, {Lin}, \&
  {Papaloizou}}]{Aarseth_etal1988}
{Aarseth}, S.~J., {Lin}, D.~N.~C., \& {Papaloizou}, J.~C.~B. 1988, ApJ, 324,
  288

\bibitem[{{Allison} et~al.(2009){Allison}, {Goodwin}, {Parker}, {de Grijs},
  {Portegies Zwart}, \& {Kouwenhoven}}]{Allison_etal2009}
{Allison}, R.~J., {Goodwin}, S.~P., {Parker}, R.~J., {de Grijs}, R., {Portegies
  Zwart}, S.~F., \& {Kouwenhoven}, M.~B.~N. 2009, ApJL, 700, L99.
  \eprint{0906.4806}

\bibitem[{{Chernoff} \& {Weinberg}(1990)}]{ChernoffWeinberg1990}
{Chernoff}, D.~F., \& {Weinberg}, M.~D. 1990, ApJ, 351, 121

\bibitem[{{Eggleton} et~al.(1989){Eggleton}, {Tout}, \&
  {Fitchett}}]{Eggleton_etal1989}
{Eggleton}, P.~P., {Tout}, C.~A., \& {Fitchett}, M.~J. 1989, ApJ, 347, 998

\bibitem[{{Fregeau} \& {Rasio}(2007)}]{FregeauRasio2007}
{Fregeau}, J.~M., \& {Rasio}, F.~A. 2007, ApJ, 658, 1047.
  \eprint{arXiv:astro-ph/0608261}

\bibitem[{{Giersz} et~al.(2008){Giersz}, {Heggie}, \&
  {Hurley}}]{Giersz_etal2008}
{Giersz}, M., {Heggie}, D.~C., \& {Hurley}, J.~R. 2008, MNRAS, 388, 429.
  \eprint{0801.3968}

\bibitem[{{Hurley} et~al.(2000){Hurley}, {Pols}, \& {Tout}}]{Hurley_etal2000}
{Hurley}, J.~R., {Pols}, O.~R., \& {Tout}, C.~A. 2000, MNRAS, 315, 543.
  \eprint{arXiv:astro-ph/0001295}

\bibitem[{{King}(1966)}]{King1966}
{King}, I.~R. 1966, AJ, 71, 64

\bibitem[{{Kroupa}(2002)}]{Kroupa2002}
{Kroupa}, P. 2002, Science, 295, 82. \eprint{arXiv:astro-ph/0201098}

\bibitem[{{Makino} \& {Aarseth}(1992)}]{MakinoAarseth1992}
{Makino}, J., \& {Aarseth}, S.~J. 1992, PASJ, 44, 141

\bibitem[{{McMillan} et~al.(2007){McMillan}, {Vesperini}, \& {Portegies
  Zwart}}]{McMillan_etal2007}
{McMillan}, S.~L.~W., {Vesperini}, E., \& {Portegies Zwart}, S.~F. 2007, ApJL,
  655, L45. \eprint{arXiv:astro-ph/0609515}

\bibitem[{{Portegies Zwart} et~al.(2009){Portegies Zwart}, {McMillan},
  {Harfst}, {Groen}, {Fujii}, {Nuall{\'a}in}, {Glebbeek}, {Heggie}, {Lombardi},
  {Hut}, {Angelou}, {Banerjee}, {Belkus}, {Fragos}, {Fregeau}, {Gaburov},
  {Izzard}, {Juri{\'c}}, {Justham}, {Sottoriva}, {Teuben}, {van Bever},
  {Yaron}, \& {Zemp}}]{PortegiesZwart_etal2009}
{Portegies Zwart}, S., {McMillan}, S., {Harfst}, S., {Groen}, D., {Fujii}, M.,
  {Nuall{\'a}in}, B.~{\'O}., {Glebbeek}, E., {Heggie}, D., {Lombardi}, J.,
  {Hut}, P., {Angelou}, V., {Banerjee}, S., {Belkus}, H., {Fragos}, T.,
  {Fregeau}, J., {Gaburov}, E., {Izzard}, R., {Juri{\'c}}, M., {Justham}, S.,
  {Sottoriva}, A., {Teuben}, P., {van Bever}, J., {Yaron}, O., \& {Zemp}, M.
  2009, New Astronomy, 14, 369. \eprint{0807.1996}

\bibitem[{{Portegies Zwart} et~al.(2001){Portegies Zwart}, {McMillan}, {Hut},
  \& {Makino}}]{PortegiesZwart_etal2001}
{Portegies Zwart}, S.~F., {McMillan}, S.~L.~W., {Hut}, P., \& {Makino}, J.
  2001, MNRAS, 321, 199. \eprint{arXiv:astro-ph/0005248}

\bibitem[{{Takahashi} \& {Portegies Zwart}(2000)}]{TakahashiPortegiesZwart2000}
{Takahashi}, K., \& {Portegies Zwart}, S.~F. 2000, ApJ, 535, 759.
  \eprint{arXiv:astro-ph/9903366}

\bibitem[{{Tanikawa} \& {Fukushige}(2009)}]{TanikawaFukushige2009}
{Tanikawa}, A., \& {Fukushige}, T. 2009, \pasj, 61, 721. \eprint{1005.2237}

\bibitem[{{Vesperini} et~al.(2006){Vesperini}, {McMillan}, \& {Portegies
  Zwart}}]{Vesperini_etal2006}
{Vesperini}, E., {McMillan}, S.~L.~W., \& {Portegies Zwart}, S.~F. 2006, in
  Joint Discussion 14, 26th IAU General Assembly

\bibitem[{{Vesperini} et~al.(2009){Vesperini}, {McMillan}, \& {Portegies
  Zwart}}]{Vesperini_etal2009}
--- 2009, in Globular Clusters - Guides to Galaxies, ESO Astrophysics Symposia,
  edited by {Richtler, T.~\& Larsen, S.} (Springer Berlin Heidelberg), p. 429

\end{thebibliography}

\end{document}